\documentclass[aps,prc,twocolumn,superscriptaddress,preprintnumbers,amsmath,amssymb,floatfix,showpacs,showkeys,nofootinbib]{revtex4}

\usepackage{graphicx, fink}
\usepackage{epsfig}
\usepackage{dcolumn}
\usepackage{bm}
\usepackage{longtable}
\usepackage{color}
\usepackage{float}
\usepackage{comment}
\usepackage{amsmath}
\usepackage{amsfonts}

\begin{document}

\newcommand{\nuc}[2]{\ensuremath{^{#1}}#2}
\newcommand{\etal}{{\em et~al.}}
\newcommand{\geant} {{\bf {G}\texttt{\scriptsize{EANT}}4}}

\preprint{MAX-lab $^{12}$C$(\gamma,\gamma)$ Summary Article for arXiv}

\title{Compton scattering from $\bm{^{12}}$C using tagged photons in the energy range 65--115 MeV}

\author{\mbox{L.~S.~Myers}}
  \altaffiliation{Present address: Thomas Jefferson National Accelerator Facility, Newport News, VA 23606, USA}
\affiliation{Department of Physics, University of Illinois at Urbana-Champaign, Urbana, IL 61801, USA}
\author{\mbox{K.~Shoniyozov}}
\affiliation{Department of Physics and Astronomy, University of Kentucky, Lexington, KY 40506, USA}
\author{\mbox{M.~F.~Preston}}
\affiliation{Department of Physics, Lund University, SE-221 00 Lund, Sweden}
\author{\mbox{M.~D.~Anderson}}
\affiliation{School of Physics and Astronomy, University of Glasgow, Glasgow G12 8QQ, Scotland UK}
\author{\mbox{J.~R.~M.~Annand}}
\affiliation{School of Physics and Astronomy, University of Glasgow, Glasgow G12 8QQ, Scotland UK}
\author{\mbox{M.~Boselli}}
  \altaffiliation{Present address: European Center for Theoretical studies in Nuclear Physics and Related Areas, Strada delle Tabarelle 286, I-38123 Villazzano, Trento, Italy}
\affiliation{School of Physics and Astronomy, University of Glasgow, Glasgow G12 8QQ, Scotland UK}
\author{\mbox{W.~J.~Briscoe}}
\affiliation{Department of Physics, The George Washington University, Washington, DC 20052, USA}
\author{\mbox{J.~Brudvik}}
\affiliation{MAX IV Laboratory, Lund University, SE-221 00 Lund, Sweden}
\author{\mbox{J.~I.~Capone}}
  \altaffiliation{Present address: Department of Physics, University of Maryland, College Park, MD 20742, USA}
\affiliation{Department of Physics, The George Washington University, Washington, DC 20052, USA}
\author{\mbox{G.~Feldman}}
\affiliation{Department of Physics, The George Washington University, Washington, DC 20052, USA}
\author{\mbox{K.~G.~Fissum}}
  \altaffiliation{Corresponding author; \texttt{kevin.fissum@nuclear.lu.se}}
\affiliation{Department of Physics, Lund University, SE-221 00 Lund, Sweden}
\author{\mbox{K.~Hansen}}
\affiliation{MAX IV Laboratory, Lund University, SE-221 00 Lund, Sweden}
\author{\mbox{S.~S.~Henshaw}}
  \altaffiliation{Present address: National Security Technologies, Andrews AFB, MD 20762, USA}
\affiliation{Department of Physics, Duke University, Durham, NC 27708, USA}
\author{\mbox{L.~Isaksson}}
\affiliation{MAX IV Laboratory, Lund University, SE-221 00 Lund, Sweden}
\author{\mbox{R.~Jebali}}
\affiliation{School of Physics and Astronomy, University of Glasgow, Glasgow G12 8QQ, Scotland UK}
\author{\mbox{M.~A.~Kovash}}
\affiliation{Department of Physics and Astronomy, University of Kentucky, Lexington, KY 40506, USA}
\author{\mbox{K.~Lewis}}
  \altaffiliation{Present Address: Trident Training Facility, Navy Submarine Base, Bangor, WA 98383, USA}
\affiliation{Department of Physics, The George Washington University, Washington, DC 20052, USA}
\author{\mbox{M.~Lundin}}
\affiliation{MAX IV Laboratory, Lund University, SE-221 00 Lund, Sweden}
\author{\mbox{I.~J.~D.~MacGregor}}
\affiliation{School of Physics and Astronomy, University of Glasgow, Glasgow G12 8QQ, Scotland UK}
\author{\mbox{D.~G.~Middleton}}
  \altaffiliation{Present address: Institut f{\"u}r Kernphysik, University of Mainz, D-55128 Mainz, Germany}
\affiliation{Kepler Centre for Astro and Particle Physics, Physikalisches Institut, Universit{\"a}t T{\"u}bingen, D-72076 T{\"u}bingen, Germany}
\author{\mbox{D.~E.~Mittelberger}}
  \altaffiliation{Present Address: LOASIS Program, Lawrence Berkeley National Lab, Berkeley, CA 94720, USA}
\affiliation{Department of Physics, The George Washington University, Washington, DC 20052, USA}
\author{\mbox{M.~Murray}}
\affiliation{School of Physics and Astronomy, University of Glasgow, Glasgow G12 8QQ, Scotland UK}
\author{\mbox{A.~M.~Nathan}}
\affiliation{Department of Physics, University of Illinois at Urbana-Champaign, Urbana, IL 61801, USA}
\author{\mbox{S.~Nutbeam}}
\affiliation{School of Physics and Astronomy, University of Glasgow, Glasgow G12 8QQ, Scotland UK}
\author{\mbox{G.~V.~O'Rielly}}
\affiliation{Department of Physics, University of Massachusetts Dartmouth, Dartmouth, MA 02747, USA}
\author{\mbox{B.~Schr\"oder}}
  \affiliation{Department of Physics, Lund University, SE-221 00 Lund, Sweden}
\affiliation{MAX IV Laboratory, Lund University, SE-221 00 Lund, Sweden}
\author{\mbox{B.~Seitz}}
\affiliation{School of Physics and Astronomy, University of Glasgow, Glasgow G12 8QQ, Scotland UK}
\author{\mbox{S.~C.~Stave}}
  \altaffiliation{Present address: Pacific Northwest National Laboratory, Richland, WA 99352, USA}
\affiliation{Department of Physics, Duke University, Durham, NC 27708, USA}
\author{\mbox{H.~R.~Weller}}
\affiliation{Department of Physics, Duke University, Durham, NC 27708, USA}

\collaboration{The COMPTON@MAX-lab Collaboration}
\noaffiliation

\date{\today}

\begin{abstract}

Elastic scattering of photons from $^{12}$C has been investigated using 
quasi-monoenergetic tagged photons with energies in the range 65 -- 115 MeV 
at laboratory angles of 60$^\circ$, 120$^\circ$, and 150$^\circ$
at the Tagged-Photon Facility at the MAX IV Laboratory in Lund, Sweden. A 
phenomenological model was employed to provide an estimate of the sensitivity 
of the \nuc{12}{C}($\gamma$,$\gamma$)\nuc{12}{C} cross section to the 
bound-nucleon polarizabilities. 

\keywords{Compton Scattering; Carbon; Polarizabilities.}

\end{abstract}

\pacs{25.20.Dc}

\maketitle

\section{\label{section:introduction}Introduction}

Much effort has been devoted to studying $\alpha$ and $\beta$, the 
electromagnetic polarizabilities of the proton and neutron. These 
polarizabilities represent the first-order responses of the internal structure 
of the nucleon to an external electric or magnetic field. 
The majority of nucleon-polarizability measurements have utilized the process
of nuclear Compton scattering. A review of these experiments can be found in 
Ref.~\cite{griesshammer2012}.

The most recent global fit \cite{griesshammer2012} to all the data up to 
170~MeV has yielded polarizabilities for the proton in units of 10$^{-4}$ 
fm$^3$ of
\begin{equation}
\label{eq:AlphaBetaP}
\begin{split}
&\alpha_p = 10.7 \pm 0.3_{\rm stat} \pm 0.2_{\rm BSR} \pm 0.8_{\rm th} \\
&\beta_p = 3.1 \mp 0.3_{\rm stat} \pm 0.2_{\rm BSR} \pm 0.8_{\rm th}, \\
\end{split}
\end{equation}

\noindent where the first uncertainty is statistical, the second is due to 
uncertainties in the Baldin Sum Rule (BSR), and the third is due to theoretical
uncertainty. The Baldin Sum Rule is given by \cite{baldin1960}
\begin{equation}
\label{eq:BaldinSR}
\alpha + \beta = \frac{1}{2\pi^2} \int_{\omega_{\rm th}}^\infty \frac{\sigma_\gamma(\omega) d\omega}{\omega^2},
\end{equation}

\noindent where $\sigma_\gamma(\omega)$ is the total photoabsorption cross 
section for the nucleon and $\omega_{\rm th}$ is the threshold energy for pion 
photoproduction.  These results were obtained under the constraint of the 
present-day evaluation \cite{deleon2001} of the BSR for the proton 
which is 
\begin{equation}
\label{eq:BSRP}
\alpha_p + \beta_p = 13.8 \pm 0.4.
\end{equation}

Similarly, the neutron polarizabilities have been extracted from 
measurements of \nuc{2}{H}($\gamma$,$\gamma$)\nuc{2}{H}.  They have been 
determined to be
\begin{equation}
\label{eq:AlphaBetaN}
\begin{split}
&\alpha_n = 11.1 \pm 1.8_{\rm stat} \pm 0.4_{\rm BSR} \pm 0.8_{\rm th} \\
&\beta_n = 4.1 \mp 1.8_{\rm stat} \pm 0.4_{\rm BSR} \pm 0.8_{\rm th} \\
\end{split}
\end{equation}

\noindent preserving the BSR for the neutron \cite{levchuk2000}
\begin{equation}\label{eq:BSRN}
\alpha_n + \beta_n = 15.2 \pm 0.4.
\end{equation}

It is also reasonable to ask whether the nucleon polarizabilities are modified when 
the proton or neutron is bound in a nucleus and, if so, to what degree. A multitude 
of Compton-scattering experiments have been carried out with a variety of light nuclei 
(see Table~\ref{table:nuclei}) for the purpose of determining the bound-nucleon 
polarizabilities ($\alpha_{\rm eff}$ and $\beta_{\rm eff}$) given by 

\begin{table}
\caption{\label{table:nuclei}
Summary of nuclei studied using Compton scattering below the energy
threshold for pion production.}
\begin{ruledtabular}
\begin{tabular}{cl}
Nucleus & Reference \\
$^4$He & \cite{wells1990}, \cite{fuhrberg1995}, \cite{proff1999}\\
$^6$Li & \cite{myers2012}\\
$^{12}$C & \cite{wright1985}, \cite{schelhaas1990}, \cite{ludwig1992}, \cite{hager1995}, \cite{warkentin2001}\\
$^{16}$O & \cite{proff1999}, \cite{myers2012}, \cite{ludwig1992}, \cite{hager1995}, \cite{mellendorf1993}, \cite{feldman1996}\\
$^{40}$Ca & \cite{proff1999}, \cite{wright1985}\\
\end{tabular}
\end{ruledtabular}
\end{table}

\begin{equation}\label{eq:bound_ab}
\alpha_{\rm eff} = \alpha_N + \Delta \alpha,~~ \beta_{\rm eff} = \beta_N + \Delta \beta,
\end{equation}

\noindent where $\alpha_N$ and $\beta_N$ are the nucleon-averaged free polarizabilities 
and $\Delta\alpha$ and $\Delta\beta$ represent the nuclear modifications \cite{warkentin2001}
which can be extracted from the scattering data.

These data sets have been analyzed using a model that parametrizes the Compton-scattering 
amplitude in terms of the photoabsorption cross section, its multipole decomposition and 
the bound-nucleon polarizabilities. The results typically produce a value for 
$\alpha_{\rm eff}$+$\beta_{\rm eff}$ that is in agreement with the free-nucleon sum rules 
given above~\cite{hutt2000}. However, though the sum is unchanged, several measurements 
\cite{fuhrberg1995,feldman1996} have reported a significant modification to the electric 
polarizability ($\Delta\alpha$ approximately $-$5 to $-$10) whereas other groups report 
bound and free polarizabilities that are nearly equal \cite{hager1995}. In this paper we 
present a substantial new data set for Compton scattering from \nuc{12}C and report on 
the extracted values of the bound-nucleon polarizabilities.

\section{Phenomenological Model}

A phenomenological model has been used to evaluate the sensitivity of the \nuc{12}{C} 
Compton-scattering data to the magnitude of the electromagnetic polarizabilities. This 
model is based on the work presented in \cite{wright1985}. The Compton-scattering amplitude 
can be written~\cite{feldman1996} in terms of the one- and two-body seagull (SG) amplitudes 
(which are explicitly dependent on the polarizabilities $\alpha_{\rm eff}$ and 
$\beta_{\rm eff}$) as
\begin{equation}\label{eq:FeldmanEq}
\begin{split}
R(E,\theta) = &R^{GR}(E,\theta) + R^{QD}(E,\theta) \\
&+ R^{SG}_1(E,\theta) + R^{SG}_2(E,\theta). \\
\end{split}
\end{equation}

The first two terms in Eq. \ref{eq:FeldmanEq} are related to the giant resonances 
($E1$, $E2$, and $M1$; hereafter refered to as GR) and the quasideuteron (QD) processes, 
respectively. The amplitudes are given by
\begin{equation}\label{eq:Amp1}
\begin{split}
R^{GR}(E,\theta) = & f_{E1}(E)g_{E1}(\theta) + f_{E2}(E)g_{E2}(\theta) \\
&+ f_{M1}(E)g_{M1}(\theta) + \frac{NZ}{A}r_0[1+\kappa_{GR}]g_{E1}(\theta),
\end{split}
\end{equation}

\noindent and
\begin{equation}\label{eq:Amp2}
\begin{split}
R^{QD}(E,\theta) = &\left[ f_{QD}(E) + \frac{NZ}{A}r_0\kappa_{QD} \right] \\
&\times F_2(q)g_{E1}(\theta), \\
\end{split}
\end{equation}

\noindent where the complex forward-scattering amplitudes are denoted by $f_\lambda(E)$ 
($\lambda$ = $E1$, $E2$, $M1$), the appropriate angular factor is $g_\lambda(\theta)$ 
(see Ref.~\cite{myers2012} for the angular factors), $r_0$ is the classical nucleon 
radius, and the enhancement factors [$1+\kappa_{GR}$] and $\kappa_{QD}$ are the integrals 
of the GR and QD photoabsorption cross sections in units of the classical dipole sum 
rule.  Since the QD process is modeled as an interaction with a neutron-proton pair, it 
is modulated by a two-body form factor $F_2(q)$ where $q$ is the momentum transfer.

The seagull amplitudes account for subnucleon and meson-exchange degrees of freedom 
and are necessary to preserve gauge invariance in the total scattering amplitude. 
The one-body seagull amplitude is
\begin{equation}\label{eq:Amp3}
\begin{split}
  R^{SG}_1(E,\theta) = & \left\{ \left[ -Zr_0 + \left( \frac{E}{\hbar c} \right)^2 A\alpha_{\rm eff} \right] g_{E1}(\theta) \right.\\
  & \left. + \left[ \left( \frac{E}{\hbar c} \right)^2 A\beta_{\rm eff} \right] g_{M1}(\theta) \right\} F_1(q),
\end{split}
\end{equation}

\noindent where the higher-order terms have been omitted. This process is modulated by 
the one-body form factor $F_1(q)$ which is given by
\begin{equation}
F_1(q) = \frac{4\pi}{q} \int_0^\infty \rho(r) \mathrm{sin}(qr) r dr ,
\end{equation}
\noindent where $\rho(r)$ is given by the three-parameter Fermi function \cite{dejager1974}
\begin{equation}
\rho(r) = \rho_0 \frac{1 + \frac{wr^2}{c^2}}{1 + \mathrm{e}^{\frac{r-c}{z}}},
\end{equation}
with $w=-$0.149, $c=$2.355 fm, $z=$ 0.522 fm, and the form factor is normalized 
so that $F_1(0)=$1.

The two-body seagull amplitude is
\begin{equation}\label{eq:Amp4}
\begin{split}
  R^{SG}_2(E,\theta) = & \left\{ \left [ -\frac{NZ}{A}\kappa r_0 + \left ( \frac{E}{\hbar c} \right )^2 A \alpha_{\rm ex} \right ]  g_{E1}(\theta) \right.\\
  & \left. + \left[ \left( \frac{E}{\hbar c} \right)^2 A\beta_{\rm ex} \right] g_{M1}(\theta) \right\} F_2(q), \\
  \end{split}
\end{equation}

\noindent where the exchange polarizabilities are denoted by $\alpha_{\rm ex}$ and 
$\beta_{\rm ex}$, $\kappa$ = $\kappa_{GR}$ + $\kappa_{QD}$, and the higher-order terms 
have been dropped. The two-body form factor is chosen by convention as 
$F_2(q)$ = $[F_1(q/2)]^2$.

The parametrization of the $E1$ resonance is taken from Ref.~\cite{wright1985} 
where the angle-averaged differential cross section was used to extract the $E1$ 
resonance below 40 MeV. The $E2$ strength between 25 and 35 MeV \cite{hager1995} and 
an $M1$ resonance \cite{hayward1977} were also included. These included resonances are 
listed in Table~\ref{table:E1_E2}.

\begin{table}
\caption{\label{table:E1_E2}
$E2$, $M1$, and QD parameters. }
\begin{ruledtabular}
\begin{tabular}{cccc}
Resonance & E$_\lambda$ (MeV) & $\sigma_\lambda$ (mb) & $\Gamma_\lambda$ (MeV) \\
\hline
$E2$ & 26.0 & 1.8 & 0.50 \\
 & 32.3 & 1.2 & 2.60 \\
 & & & \\
$M1$ & 15.1 & 29780 & 37$\times$10$^{-6}$ \\
 & & & \\
QD & 40 & 1.0 & 100 \\
\end{tabular}
\end{ruledtabular}
\end{table}

Levinger's modified quasideuteron model \cite{levinger79} and a damped Lorentzian 
lineshape were used to define a piecewise function to parametrize the QD process
(Eq.~\ref{eq:qd_para1}). This parametrization was fitted to the existing total 
photoabsorption cross-section data~\cite{ahrens75} above 50~MeV in order to 
establish the normalization. The QD scattering cross section was taken to be
\begin{equation}\label{eq:qd_para1}
  \sigma_{QD}(E) = \left\{
    \begin{array}{lr}
      \frac{1}{2}\left [1+\tanh \frac{E-E_{t}}{\Delta E} \right ] \mathcal{L}_{QD}(E) & :E < {\rm 50~MeV}
      \\[12pt]
      L e^{-D/E} \left[\frac{NZ}{A}\right] \sigma_{D}(E) & :E > {\rm 50~MeV},\\
    \end{array}
  \right.
\end{equation}

\noindent where $\sigma_{D}(E)$ is the deuteron photoabsorption cross section 
and the parameters $L=$ 5.0 and $D=$ 5.4 were determined from the fit to the data. 
The Lorentzian $\mathcal{L}_{QD}(E)$ has the parameters $E_{QD}$, $\Gamma_{QD}$, 
and $\sigma_{QD}$ given in Table~\ref{table:E1_E2}, with $E_{t}=$ 40 MeV and 
$\Delta E=$ 10 MeV. Since the analysis of Warkentin {\it et~al.} \cite{warkentin2001} 
indicated that the extraction of $\alpha_{\rm eff}$ and $\beta_{\rm eff}$ depends 
only slightly on the parametrization of the QD amplitude, only the above 
parametrization will be used in this analysis.

\section{\label{section:expt}Experiment}

\begin{figure}[h]
\begin{center}
\includegraphics[width=0.4\textwidth]{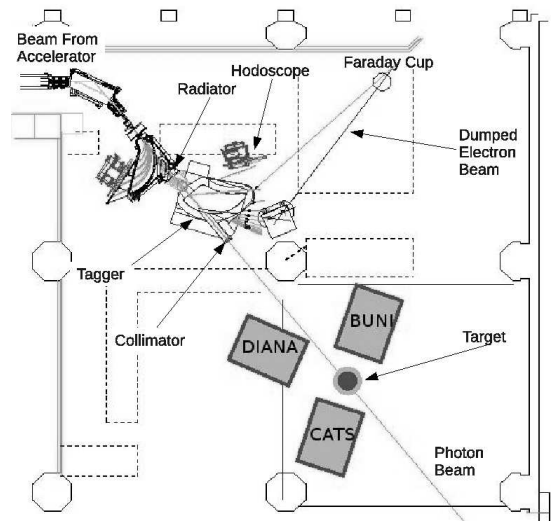}
\caption{
The layout of the experimental area showing the location of the
focal-plane hodoscope, $^{12}$C target, and NaI(Tl) detectors labeled DIANA, 
BUNI, and CATS.
\label{figure:figure_XX_setup}}
\end{center}
\end{figure}

The experiment was performed at the Tagged-Photon Facility~\cite{tpf,adler2012}
located at the MAX IV Laboratory \cite{m4} in Lund, Sweden.  A pulse-stretched
electron beam~\cite{lindgren2002} with nominal energies of 144~MeV and 165~MeV, 
a current of 15~nA, and a duty factor of 45\% was used to produce 
quasi-monoenergetic photons in the energy range 65 -- 115 MeV via the 
bremsstrahlung-tagging technique~\cite{adler1990,adler1997}. An overview of the 
experimental layout is shown in Fig.~\ref{figure:figure_XX_setup}.

The size of the photon beam was defined by a tapered tungsten-alloy primary 
collimator of 19~mm nominal diameter.  The primary collimator was followed 
by a dipole magnet and a post-collimator which were used to remove any charged 
particles produced in the primary collimator. The beam spot at the target 
location was approximately 60~mm in diameter.

The tagging efficiency~\cite{adler1997} is the ratio of the number of tagged
photons which struck the target to the number of post-bremsstrahlung electrons 
which were registered by the associated focal-plane channel.  It was measured 
absolutely during the experiment startup with three large-volume NaI(Tl) photon 
spectrometers placed directly in the beam (see below) and it was monitored 
during the experiment itself on a daily basis using a lead-glass photon detector. 
The tagging efficiency was determined to be (44 $\pm$ 1)\% throughout the experiment.

A graphite block 5.22 cm thick was used as a target. The density of the target 
was measured to be (1.83$\pm$0.02)~g/cm$^3$. The target was positioned such 
that the photon beam was perpendicular to the face of the target resulting in 
a target thickness of (4.80$\pm$0.07) $\times$ 10$^{23}$ nuclei/cm$^2$. The 
average loss of incident photon-beam flux due to absorption in the target was
approximately 7\%.

Three large-volume, segmented NaI(Tl) detectors labeled BUNI~\cite{miller1988}, 
CATS~\cite{wissmann1994}, and DIANA~\cite{myers2010} in 
Fig.~\ref{figure:figure_XX_setup} were used to detect the Compton-scattered 
photons. The detectors were located at laboratory angles of 60$^\circ$, 120$^\circ$, 
and 150$^\circ$. These detectors were each composed of a single, large NaI(Tl) 
crystal surrounded by optically-isolated, annular NaI(Tl) segments.  The detectors 
have an energy resolution of better than 2\% at energies near 100 MeV. Such 
resolution is necessary to unambiguously separate elastically scattered photons 
from those originating from the breakup of deuterium, a parallel and ongoing 
experimental effort to be reported upon in the near future.

\section{\label{section:data_analysis}Data Analysis}

\subsection{\label{subsection:yield_extraction}Yield Extraction}

\begin{figure}[h]
\begin{center}
\resizebox{0.4\textwidth}{!}{\includegraphics{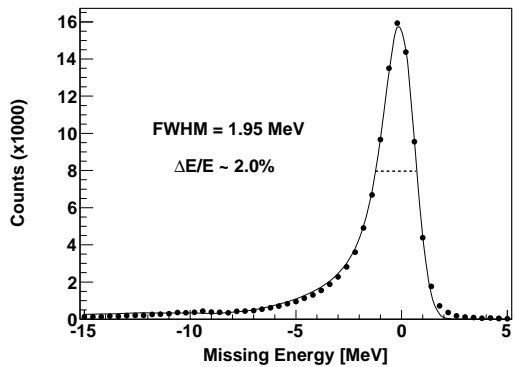}}
\caption{
Typical in-beam detector response to incident photons as a function of missing 
energy. The curve is the simulated \geant\ detector response fitted to the data.
\label{figure:figure_XX_inbeam}}
\end{center}
\end{figure}

The signals from each detector were passed to analog-to-digital converters 
(ADCs) and time-to-digital converters (TDCs) and the data recorded on an
event-by-event basis. The comprehensive dataset 
presented in this paper was acquired over a five-year period from 2007 to 
2012. During this time, the single-hit TDCs used to instrument the tagger 
focal plane were complemented with multi-hit TDCs.  Data obtained using both 
types of TDCs are presented here~\footnote{
The interested reader is directed 
to Refs.~\cite{myers2013,preston2013} for a detailed discussion and comparison 
of the results obtained using the two different types of TDCs.}.
The ADCs allowed reconstruction of the scattered-photon energies, while the 
TDCs enabled coincident timing between the NaI(Tl) detectors and the focal-plane 
hodoscope.  The energy calibration of each detector was determined by placing 
them directly into the photon beam and observing their response as a function 
of tagged-photon energy. A typical measured in-beam lineshape together 
with a \geant\ simulation \cite{agostinelli2003} of the response function of 
the detector fitted to the data is shown in Fig.~\ref{figure:figure_XX_inbeam}.

\begin{figure}[h]
\begin{center}
\resizebox{0.4\textwidth}{!}{\includegraphics{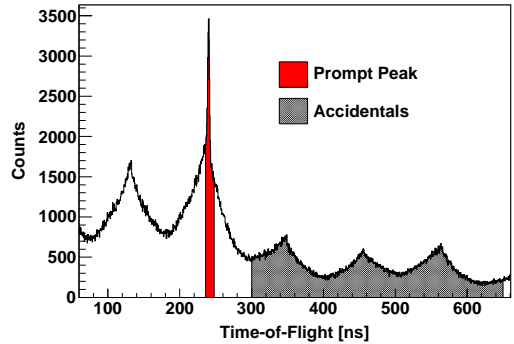}}
\caption{(Color online) The focal-plane TDC spectrum for the scattering data. 
The prompt (red) and the accidental (grey) windows are indicated.
\label{figure:figure_XX_tdc}}
\end{center}
\end{figure}

Large backgrounds arose when the detectors were moved to the various scattering 
angles and the beam intensity was increased from 10-100 Hz (for in-beam runs) 
to 1-4 MHz.  Untagged bremsstrahlung photons (related to the beam intensity) and 
cosmic rays (constant) were the dominant sources of background. An energy cut 
that accepted only events in the tagged-energy range enabled the prompt peak 
(representing coincidences between electrons in the focal-plane hodoscope and 
events in the NaI(Tl) detectors) to be identified in the focal-plane TDC spectra 
(see Fig.~\ref{figure:figure_XX_tdc}). For each NaI(Tl) detector, events occurring 
within the prompt peak were selected and a prompt missing-energy (ME) spectrum
was filled. ME was defined as the difference between the detected photon energy 
and the expected photon energy based upon the tagged-electron energy. A second 
cut was placed on an accidental (or random) timing region and an accidental ME 
spectrum was filled. This process was carried out for each focal-plane channel.

\begin{figure}[h]
\begin{center}
\resizebox{0.4\textwidth}{!}{\includegraphics{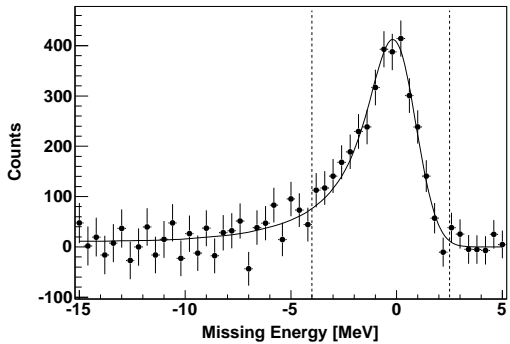}}
\caption{A typical missing-energy spectrum for a focal-plane bin obtained 
in one of the scattering configurations, with both cosmics and accidentals 
removed.  The solid line is the fit of the \geant\ lineshape to the data. The 
dashed lines indicate the ROI used to determine the yield of Compton-scattered
photons (see text).
\label{figure:figure_XX_scatter_lineshape}}
\end{center}
\end{figure}

A net sum ME spectrum for each focal-plane channel was generated by removing 
both the cosmic-ray and untagged-photon backgrounds. Due to the complex nature 
of the time structure that exists in the focal-plane TDC spectrum, 
this process was carried out in two steps. First, the cosmic-ray contribution
was subtracted from both the prompt and accidental ME spectra by 
normalizing these spectra in the energy region above the electron-beam energy. 
Next, the cosmic-subtracted accidentals were removed from the cosmic-subtracted 
prompts by normalizing the two spectra in the energy range above the 
tagged-photon energy corresponding to the particular focal-plane channel but 
below the electron-beam energy. The focal plane was divided into 4 energy bins, 
each approximately 9~MeV wide. The background-corrected ME spectrum for each 
tagged channel in a particular energy bin was then summed to create a ME spectrum 
for that bin, such as the one shown in Fig.~\ref{figure:figure_XX_scatter_lineshape}.

A \geant\ simulation was employed to determine the total yield in the 
elastic-scattering peak and 
also to quantify any corrections due to finite geometrical effects. The simulation 
output was first determined for the case of a NaI(Tl) detector positioned directly 
in the low-intensity photon beam ($\theta$=0$^{\circ}$) as shown in 
Fig.~\ref{figure:figure_XX_inbeam}.  This intrinsic simulation was then smeared with a 
Gaussian function to phenomenologically
account for the individual characteristics of each NaI(Tl) detector that are 
difficult to model in \geant. The simulated detector response, with the smearing 
determined as above, was then fitted to the scattering data over the region of 
interest (ROI) indicated by the vertical dashed lines shown in 
Fig.~\ref{figure:figure_XX_scatter_lineshape}. The fitted \geant\ lineshape 
was then used to correct for the detection efficiency of the NaI(Tl) detector 
in the ROI. This efficiency accounts for events that deposit some energy 
outside the ROI in the detector. Additionally, the correction factor for photons 
absorbed by the target and the correction to the detector acceptance due to 
the finite geometry of the experimental setup were obtained from this simulation.

\subsection{\label{subsection:normalization}Normalization}

The scattering-photon yield was then normalized to the number of photons incident
on the target and corrected for rate-dependent factors. The number of photons 
incident on the target was determined from the number of post-bremsstrahlung 
electrons detected in each focal-plane channel and the measured tagging efficiency. 
The rate-dependent corrections included ``stolen'' trues \cite{owens1990}, 
``missed'' trues, and ``ghost'' events \cite{macgibbon1995_1}. A stolen true arose 
when a random electron was detected in the focal-plane channel prior to the 
electron corresponding to the tagged photon. This correction was only applied to 
the single-hit TDC data. It was determined using the method outlined in 
Ref.~\cite{vanhoorebeke1992} and was typically 20 -- 45\%. Missed trues resulted 
from deadtime effects in the focal-plane instrumentation electronics.  Ghost events 
were an artifact of the physical overlap of the focal-plane counters. The 
missed-trues and ghost corrections were determined using a Monte Carlo simulation 
of the focal-plane electronics and amounted to approximately 5\% and 1\%, 
respectively\footnote{The interested reader is directed to Ref.~\cite{myers2013} 
for a detailed discussion of the focal-plane simulation.}.

\subsection{\label{subsubsection:systematic_uncertainities}Systematic Uncertainties}

The systematic uncertainties in this experiment were grouped into three types. 
The first was an overall scale systematic uncertainty that affected the data 
obtained at all angles and energies equally. This uncertainty arose from 
normalization factors such as the tagging efficiency.  The second type of uncertainty 
varied only with angle but not energy and was due to the acceptance of the individual 
NaI(Tl) detectors. This uncertainty had two origins: (1) the distance and aperture 
size of the detector in its scattering location; and (2) the effect of placing cuts 
on the data during analysis. Finally, certain uncertainties were strongly dependent 
on kinematics and varied with both energy and angle such as the stolen-trues 
correction. The dominant sources of systematic uncertainties are listed in 
Table~\ref{table:systs} along with typical values. The systematic uncertainties 
were combined in quadrature to obtain the overall systematic uncertainty.

\begin{table}
\caption{\label{table:systs}
Systematic uncertainties.}
\begin{ruledtabular}
\begin{tabular}{llr}
          Type &            Variable &     Value \\
\hline
         Scale &  Tagging Efficiency & $\sim$1\% \\
               &    Target Thickness & $\sim$1\% \\
               &        Missed Trues & $\sim$1\% \\
               &        Ghost Events & $\sim$2\% \\
\hline
       Angular & Detector Acceptance &    3--4\% \\
\hline
Point-to-Point &        Stolen Trues &    2--4\% \\
\hline
         Total &                     &    5--7\% \\
\end{tabular}
\end{ruledtabular}
\end{table}

\subsection{Cross Sections} \label{subsubsection:Cross Sections}

\begin{figure}[h]
\begin{center}
\resizebox{\columnwidth}{!}{\includegraphics{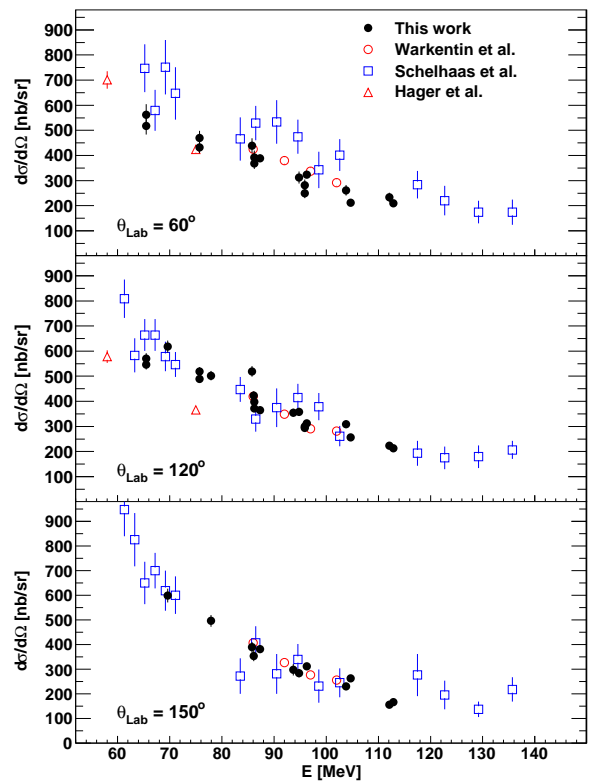}}
\caption{(Color online) Measurements of the $^{12}$C Compton-scattering cross section 
from previous experiments compared to the results from the current experiment. 
Statistical uncertainties are shown.
\label{figure:figure_XX_world_data}}
\end{center}
\end{figure}

The $^{12}$C elastic scattering cross sections measured in this experiment are presented in 
Table~\ref{table:data}. The results are also shown in 
Fig.~\ref{figure:figure_XX_world_data}, along with the results 
from \cite{schelhaas1990,ludwig1992,hager1995,warkentin2001} above 55~MeV.  
The new data are in excellent agreement with the results from 
Schelhaas~{\it et~al.}~\cite{schelhaas1990} and 
Warkentin~{\it et~al.}~\cite{warkentin2001}. 

\begin{table}
\caption{\label{table:data}
Measured cross sections for $^{12}$C($\gamma,\gamma)$ at the lab angles listed. 
The type of TDC used to record coincidences is indicated. 
The first uncertainty is statistical and the second uncertainty is total 
systematic.}
\begin{ruledtabular}
\begin{tabular}{crrr}
$E_{\gamma}$ &   $\frac{d\sigma}{d\Omega}(60^\circ)$ &   $\frac{d\sigma}{d\Omega}(120^\circ)$ &        $\frac{d\sigma}{d\Omega}(150^\circ)$ \\
       (MeV) &                               (nb/sr) &                                (nb/sr) &                                     (nb/sr) \\
\hline
\multicolumn{4}{c}{Single-hit TDCs, $E_{\rm beam}$ = 144 MeV} \\
        69.6 &                                                 &     618  $\pm$ 24  $\pm$ 42                       &                    599  $\pm$ 28  $\pm$ 32  \\
        77.9 &                                                 &     502  $\pm$ 19  $\pm$ 32                       &                    496  $\pm$ 24  $\pm$ 26  \\
        86.1 &                                                 &     423  $\pm$ 17  $\pm$ 26                       &                    354  $\pm$ 21  $\pm$ 18  \\
        93.4 &                                                 &     354  $\pm$ 16  $\pm$ 23                       &                    298  $\pm$ 23  $\pm$ 16  \\
             &                                                 &                                                   &                                             \\
\multicolumn{4}{c}{Single-hit TDCs, $E_{\rm bḅeam}$ = 165 MeV} \\
        85.8 &    439  $\pm$ 29  $\pm$ 29                      &     519  $\pm$ 21  $\pm$ 34                       &                    389  $\pm$ 21  $\pm$ 22  \\
        94.8 &    312  $\pm$ 24  $\pm$ 19                      &     358  $\pm$ 16  $\pm$ 23                       &                    284  $\pm$ 17  $\pm$ 18  \\
       103.8 &    261  $\pm$ 21  $\pm$ 17                      &     309  $\pm$ 15  $\pm$ 20                       &                    230  $\pm$ 14  $\pm$ 13  \\
       112.1 &    233  $\pm$ 19  $\pm$ 16                      &     223  $\pm$ 12  $\pm$ 15                       & 156  $\pm$ 13  $\pm$ \textcolor{white}{1}9  \\
             &                                                 &                                                   &                                             \\
\multicolumn{4}{c}{Multi-hit TDCs, $E_{\rm bḅeam}$ = 165 MeV} \\
        87.3 &    389  $\pm$ 14  $\pm$ 16                      &     365  $\pm$ 11  $\pm$ 14                       &                    381  $\pm$ 11  $\pm$ 14  \\
        96.3 &    324  $\pm$ 13  $\pm$ 13                      &     312  $\pm$ 10  $\pm$ 13                       &                    312  $\pm$ 10  $\pm$ 13  \\
       104.7 &    212  $\pm$ 11  $\pm$ 11                      &     256  $\pm$ \textcolor{white}{1}8  $\pm$ 11    & 263  $\pm$ \textcolor{white}{1}8  $\pm$ 13  \\
       112.9 &    209  $\pm$ \textcolor{white}{1}9  $\pm$ 10   &     213  $\pm$ \textcolor{white}{1}7  $\pm$ 10    & 166  $\pm$ \textcolor{white}{1}7  $\pm$ \textcolor{white}{1}8  \\
             &                                                 &                                                   &                                             \\
\multicolumn{4}{c}{Single-hit TDCs, $E_{\rm bḅeam}$ = 165 MeV} \\
        65.5 &    562  $\pm$ 42  $\pm$ 35                      &     570  $\pm$ 22  $\pm$ 29                       & \\
        75.7 &    470  $\pm$ 28  $\pm$ 29                      &     519  $\pm$ 17  $\pm$ 25                       & \\
        86.2 &    392  $\pm$ 27  $\pm$ 22                      &     398  $\pm$ 16  $\pm$ 18                       & \\
        95.9 &    281  $\pm$ 23  $\pm$ 14                      &     294  $\pm$ 15  $\pm$ 13                       & \\
             &                                                 &                                                   & \\
\multicolumn{4}{c}{Multi-hit TDCs, $E_{\rm bḅeam}$ = 165 MeV} \\
        65.5 &    517  $\pm$ 34  $\pm$ 22                      &     546  $\pm$ 19  $\pm$ 24                       & \\
        75.7 &    432  $\pm$ 23  $\pm$ 18                      &     489  $\pm$ 15  $\pm$ 20                       & \\ 
        86.2 &    368  $\pm$ 22  $\pm$ 16                      &     372  $\pm$ 14  $\pm$ 15                       & \\
        95.9 &    250  $\pm$ 20  $\pm$ 10                      &     300  $\pm$ 14  $\pm$ 12                       & \\
\end{tabular}
\end{ruledtabular}
\end{table}

\section{\label{section:results}Results}

\begin{figure}[h]
\begin{center}
\resizebox{\columnwidth}{!}{\includegraphics{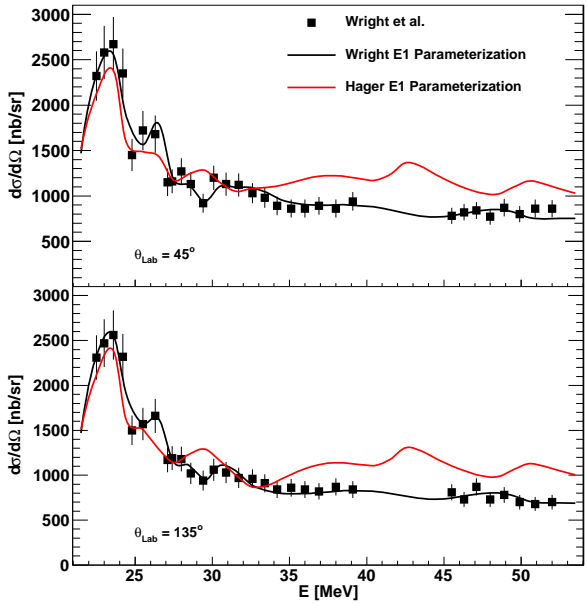}}
\caption{(Color online) Comparison of the fit to the Wright~{\it et~al.} 
data~\cite{wright1985} using the $E1$ resonance parametrization suggested by
the author (black) as well as that of H{\"a}ger {\it et al.}~\cite{hager1995} (red).
\label{figure:figure_XX_hager_vs_wright}}
\end{center}
\end{figure}

The most recent interpretations of $^{12}$C($\gamma$,$\gamma$) cross-section 
data~\cite{hager1995,warkentin2001} utilize multiple Lorentzian lineshapes to 
construct the $E1$ scattering amplitude. However, in our analysis, the 
phenomenological model is unable to fit the low-energy data of 
Wright~{\it et~al.}~\cite{wright1985} using these lineshapes 
(see Fig.~\ref{figure:figure_XX_hager_vs_wright}). 
In an attempt to incorporate all the published data, we have elected to use the 
analysis procedure detailed in~\cite{wright1985} where the $E1$ resonance is 
deduced from the low-energy ($\le$ 40~MeV), angle-averaged Compton-scattering 
cross section, and the QD scattering amplitude is given by Eq.~\ref{eq:qd_para1}. 
The $E2$ and $M1$ resonances below 40 MeV are included
for completeness, but have little effect on the results.

As suggested by Wright~{\it et~al.}~\cite{wright1985} and reinforced by 
Warkentin~{\it et~al.}~\cite{warkentin2001}, 
we also allowed for the possibility of E2 strength above 50~MeV. A third $E2$ 
resonance was added to the phenomenological model (with an equivalent lineshape 
subtracted from the QD parametrization so as to not affect the total 
photoabsorption cross section~\cite{mellendorf1993}) with a width of 30~MeV. 
The best fits to the data were achieved with an $E2$ resonance energy of 
approximately 90~MeV and a peak strength of approximately 0.2~mb. These values 
are consistent with the resonance assumed by Warkentin~{\it et~al}. The addition 
of this $E2$ resonance reduced $\chi^2$ by approximately 30\% compared to an 
identical fit without the additional $E2$ strength.

\begin{figure}[h]
\begin{center}
\resizebox{\columnwidth}{!}{\includegraphics{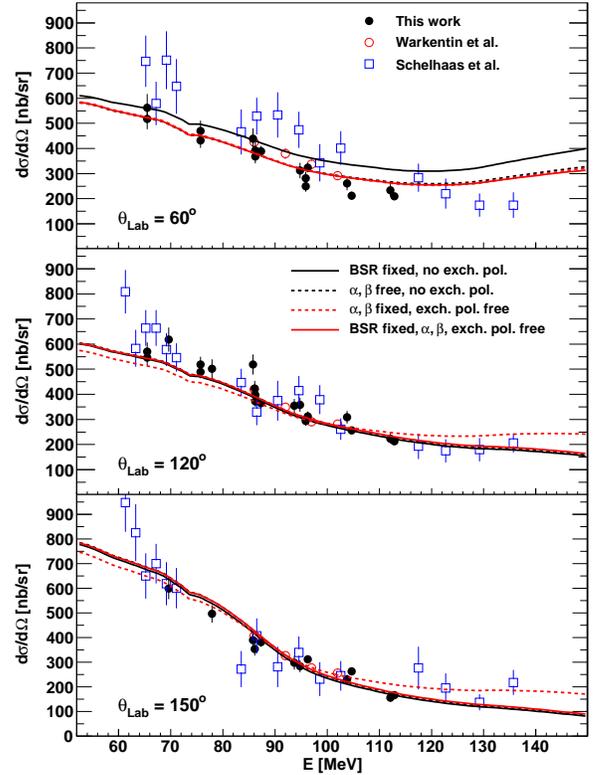}}
\caption{(Color online) Comparison of the fits to the data obtained by varying 
$\alpha_{\rm eff,ex}$ and $\beta_{\rm eff,ex}$ as described in the text. The 
uncertainties shown are the statistical and systematic uncertainties added in 
quadrature.
\label{figure:figure_XX_fits}}
\end{center}
\end{figure}

With the above parametrization, we were able to fit the entire world data set 
(excluding the data from H{\"a}ger~{\it et~al.}~\cite{hager1995})
below 150~MeV with the phenomenological model. Four different approaches were 
used involving different combinations of the BSR, the bound-nucleon polarizabilities 
($\alpha_{\rm eff}$ and $\beta_{\rm eff}$), and the exchange polarizabilities 
($\alpha_{\rm ex}$ and $\beta_{\rm ex}$) (see Table~\ref{table:fit_results}). 
In approach (1), $\alpha_{\rm eff}$ and $\beta_{\rm eff}$ were varied under the 
BSR constraint while $\alpha_{\rm ex}$ = $\beta_{\rm ex}$ = 0 were fixed. In
approach (2), the effective polarizabilities were fixed while the exchange 
polarizabilities were varied. In approach (3), $\alpha_{\rm ex}$ = $\beta_{\rm ex}$ 
= 0 were once again fixed, and $\alpha_{\rm eff}$ and $\beta_{\rm eff}$ were 
varied without the BSR constraint. In approach (4), using only the BSR constraint, 
all the polarizabilities were allowed to vary in order to minimize the $\chi^2$/DOF. 
In all four cases, the additional $E2$ resonance was fixed at E$_{\rm res}=$ 
89~MeV, $\sigma_{\rm res}=$ 0.22~mb, and 
$\Gamma_{\rm res}=$ 30~MeV) as minor variations in the resonance parameters had 
a negligible effect on the results. Together with the $E1$ parametrization 
developed by Wright~{\it et al.}~\cite{wright1985}, this analysis presents a 
consistent framework for fitting the scattering data from photon energies below 
the giant dipole resonance to energies near the threshold for pion production.  
The results are summarized in Table~\ref{table:fit_results} (quantities listed 
without uncertainties were held fixed during the fitting) and are shown in 
Fig.~\ref{figure:figure_XX_fits}. 

\begin{table}
\caption{\label{table:fit_results}
Extracted values of the effective and exchange polarizabilities subject to the
constraints outlined in the text. }
\begin{ruledtabular}
\begin{tabular}{cccccc}
approach & $\alpha_{\rm eff}$+$\beta_{\rm eff}$ & $\alpha_{\rm eff}$ & $\beta_{\rm eff}$ & $\alpha_{\rm ex}$ & $\beta_{\rm ex}$ \\
\hline
     (1) &                                 14.5 &          3.4 (0.2) &        11.1 (0.2) &                 0 &                0 \\
     (2) &                                 14.5 &               10.9 &               3.6 &        -3.9 (0.2) &        6.4 (0.2) \\
     (3) &                           18.3 (0.3) &          4.9 (0.2) &        13.4 (0.2) &                 0 &                0 \\
     (4) &                                 14.5 &          3.6 (1.1) &        10.9 (1.1) &         1.3 (0.8) &        2.1 (0.7) \\
\end{tabular}
\end{ruledtabular}
\end{table}

The extracted value of $\alpha_{\rm eff}$ varied over the range 3--11. 
Additionally, the exchange polarizabilities were quite large depending on 
the values used for $\alpha_{\rm eff}$ and $\beta_{\rm eff}$. Thus, it is 
clear that this model for Compton scattering is unable to differentiate 
between in-medium modifications to the free-nucleon polarizabilities and 
the effects of the two-body exchange polarizabilities.
The results of our analysis indicate that the net electric polarizability 
of the bound nucleon ($\alpha_{\rm eff}$+$\alpha_{\rm ex}$) is significantly 
reduced from its free value and that the magnetic polarizability is much 
larger than its free value. This is in direct opposition to the results 
reported by H{\"a}ger~{\it et al.}~\cite{hager1995} where the observed 
bound-nucleon polarizabilities were in agreement with the free values. The 
sources of this discrepancy are the reported cross sections, especially at 
the backward scattering angles (see Fig.~\ref{figure:figure_XX_world_data}), 
and the choice of the model parametrization. Fitting the 
H{\"a}ger~{\it et al.}~\cite{hager1995} data alone with the model developed 
in this paper (with $\alpha_{\rm ex}$ = $\beta_{\rm ex}$ = 0 fixed) produces 
a value of $\alpha_{\rm eff}$ = 8.2 $\pm$ 0.5 and $\beta_{\rm eff}$ = 6.3 
$\mp$ 0.5. Thus, the choice of the resonance parametrization explains part 
of the discrepancy. We note that the H{\"a}ger~{\it et al.}~\cite{hager1995} 
cross-section data, especially at the backward angles, are much smaller than 
those reported in most other 
experiments~\cite{schelhaas1990,ludwig1992,warkentin2001}. The effect of a 
smaller cross section is an increase in the difference
$\alpha_{\rm eff}-\beta_{\rm eff}$ which, in turn, produces values for the 
bound polarizabilities much closer to their free values. Based on our data 
and the analysis presented here, we assert that either the bound-nucleon 
polarizabilities differ considerably from the free-nucleon values or there 
are substantial contributions of two-body exchange polarizabilities. Both 
of these statements agree with the conclusions of 
Feldman~{\it et~al.}~\cite{feldman1996} drawn based upon 
$^{16}$O($\gamma$,$\gamma$) data.

\section{Conclusions}

In this work, we present a new measurement of the $^{12}$C Compton-scattering 
cross section for the energy range 65--115~MeV. The results are in good 
agreement with the previously published results of 
Schelhaas~{\it et~al.} \cite{schelhaas1990}, Ludwig~{\it et~al.} \cite{ludwig1992}, 
and Warkentin~{\it et~al.} \cite{warkentin2001}. However, there is a substantial 
discrepancy with the results reported by H{\"a}ger~{\it et~al.} \cite{hager1995}.

The values of the extracted bound-nucleon polarizabilities were found to be strongly 
dependent on the parametrization of the cross section. The range of extracted 
$\alpha_{\rm eff}$ was 3--11 depending on whether or not the exchange polarizabilities 
were included. Based on the results and analysis, there are in-medium effects and/or 
exchange polarizabilities that must be accounted for in a full calculation of the 
Compton-scattering process. Unfortunately, the current world-data set does not
indicate which of these effects is more important. The data do seem to have a 
strong preference for additional $E2$ strength located above 50~MeV which could 
be experimentally determined.

\begin{acknowledgments}

The authors acknowledge the outstanding support of the staff of the MAX IV Laboratory 
which made this experiment successful. We also gratefully acknowledge the Data 
Management and Software Centre, a Danish contribution to the European Spallation
Source ESS AB, for generously providing access to their computations cluster. The 
Lund group acknowledges the financial support of the Swedish Research Council, the 
Knut and Alice Wallenberg Foundation, the Crafoord Foundation, the Swedish Institute, 
the Wenner-Gren Foundation, and the Royal Swedish Academy of Sciences. This work was 
sponsored in part by the U.S. Department of Energy under grants DE-FG02-95ER40901, 
DE-FG02-99ER41110, and DE-FG02-06ER41422, the US National Science Foundation 
under Award Number 0853760 and the UK Science and Technology Facilities Council.

\end{acknowledgments}

\bibliography{myers_etal}

\end{document}